# Prediction of Drug-Induced TdP Risks Using Machine Learning and Rabbit Ventricular Wedge Assay

Jaela Foster-Burns [a], Nan Miles Xi [b,*]

[a] *Department of Biology, Loyola University Chicago, Chicago, IL 60660, USA*

[b] *Department of Mathematics and Statistics, Loyola University Chicago, Chicago, IL 60660, USA*

[*] Correspondence: Nan Miles Xi (mxi1@luc.edu)

# ABSTRACT

Torsades de pointes (TdP) is an irregular heart rhythm as a side effect of drugs and may cause sudden cardiac death. A machine learning model that can accurately identify drug TdP risk is necessary. This study uses multinomial logistic regression models to predict three-class drug TdP risks based on datasets generated from rabbit ventricular wedge assay experiments. The training-test split and five-fold cross-validation provide unbiased measurements for prediction accuracy. We utilize bootstrap to construct a 95% confidence interval for prediction accuracy. The model interpretation is further demonstrated by permutation predictor importance. Our study offers an interpretable modeling method suitable for drug TdP risk prediction. Our method can be easily generalized to broader applications of drug side effect assessment.

## INTRODUCTION

Torsades de Pointes (TdP) is a fatal polymorphic ventricular tachycardia. It is distinguished by the elevated beating of the heart's lower chambers (ventricles) and QT prolongation [1,2]. The risk of TdP may be signified on an electrocardiogram through a display of oscillatory changes in amplitude of the QRS complexes around the isoelectric line. Experimental studies and literature indicate that pharmaceutical drugs known to treat cardiomyopathies (heart disease) fall into different risk levels of inducing TdP [3]. In 2005 The International Committee/Council for Harmonisation (ICH) established international regulatory guidelines, ICH S7B and ICH E14, for the pharmaceutical industry to analyze drug TdP risk [4]. Subsequentially, in 2013 the US Food and Drug Administration (FDA) proposed the Comprehensive In Vitro Proarrhythmic Assay (CiPA) initiative for improved drug-induced TdP risk prediction [5]. A cumulation of these guidelines in combination with *in vitro* (in the test tube, not in the organism) protocols has been utilized to determine drug TdP risk within preclinical and clinical studies. These protocols use multielectrode array or voltage-sensing optical approaches, combined with logistical ordinal and multinomial linear regression models, to analyze electrophysiologic responses to drugs commonly linked to low, medium, or high TdP risk categories [2,3].

The goals of this study surround the usage of data generated from *in vitro* preclinical studies to predict drug TdP risk, along with the examination of variables contributing to drug TdP risk. This study further examines the uncertainty of TdP drug risk prediction and identifies important variables in the prediction. To achieve these goals, a multinomial logistic regression model, training-test split, cross-validation, and permutation predictor importance are used in this study. The drug TdP risk prediction is a three-classification problem that can be modeled with multinomial random variables. The data for the model was acquired from the literature on a rabbit ventricular wedge experiment [6].

The main result of this study indicates the risk classifications of 28 drugs known to induce TdP. The risk classifications are determined by *in vitro* preclinical studies and are mainly supported by ECG readings. The model places any drug into a low-, medium-, or high-risk category and reports on the accuracy of the risk prediction while giving prediction uncertainty and a 95% confidence interval. The significance of this study is that a model is trained to predict unknown data with extremely high accuracy while stating the most important predictor. This study also outlines tools to improve accuracy (cross-validation and bootstrap) or better the model.

## DATASETS

This study utilizes data pulled from the literature on the utility of a normalized TdP score system in drug proarrhythmic potential [6]. The data in this literature originates from experimentation done on rabbit ventricular wedge assays (RVWAs). The experimentation consists of electrophysiological recordings conducted on surgically prepared rabbit left ventricular wedges and is further detailed in supported literature [7]. From experimentation, a transmural-pseudo-electrocardiogram (ECG) is recorded, allowing for the calculation of TdP scores for each drug and the development of TdP risk categories – low, medium, or high risk of TdP. This study uses the findings from the electrocardiogram recordings to generate 15 variables to build the model introduced (Table 1). Four replicates on each of the 28 drugs tested with RVWAs produce 112 observations in total for the model to be built upon (Table 2).

## METHODS

The model in this study is a multinomial logistic regression model [8]. It classifies any drug into one out of three risk categories. The three classifications are represented in a response variable $Y$ and denoted by the letters $L$, $M$, and $H$. The $L$ denotes low risk of TdP, the $M$ denotes medium risk of TdP, and the $H$ denotes high risk of TdP. The response variable $Y$ is dependent upon $p$ predictors: $X_1, X_2, \ldots, X_p$, used to model probability. In the multinomial logistic regression model, it is assumed that $Y$ is a multinomial random variable, as shown in equation (1)

$$Y \sim \begin{cases} P(Y = L) = p_L \\ P(Y = M) = p_M \\ P(Y = H) = p_H = 1 - p_L - p_M \end{cases} \quad (1)$$

The probability of TdP risk can be modeled by following equations (2), (3), and (4)

$$p_L = P(Y = L) = \frac{e^{(\beta_{01}+\beta_{11}X_1+\cdots+\beta_{p1}X_p)}}{e^{(\beta_{01}+\beta_{11}X_1+\cdots+\beta_{p1}X_p)} + e^{(\beta_{02}+\beta_{12}X_1+\cdots+\beta_{p2}X_p)} + e^{(\beta_{03}+\beta_{13}X_1+\cdots+\beta_{p3}X_p)}} \quad (2)$$

$$p_M = P(Y = M) = \frac{e^{(\beta_{02}+\beta_{12}X_1+\beta_{22}X_2+\cdots+\beta_{p2}X_p)}}{e^{(\beta_{01}+\beta_{11}X_1+\cdots+\beta_{p1}X_p)} + e^{(\beta_{02}+\beta_{12}X_1+\cdots+\beta_{p2}X_p)} + e^{(\beta_{03}+\beta_{13}X_1+\cdots+\beta_{p3}X_p)}} \quad (3)$$

$$p_H = P(Y = H) = 1 - p_L - p_M \quad (4)$$

where $\beta$s are model parameters. With $p_L$, $p_M$, $p_H$, the final three-class model output is obtained by (5)

$$Y = \begin{cases} L & if\ p_L = \max(p_L, p_M, p_H) \\ M & if\ p_M = \max(p_L, p_M, p_H) \\ H & if\ p_H = \max(p_L, p_M, p_H) \end{cases} \quad (5)$$

The maximum likelihood estimation is used to obtain the model parameters, in which the model parameters maximize the probability of observing data [9]. This produces TdP risk predictions for each observation according to a three-class system (low, medium, or high risk of TdP). The model gives prediction accuracy following the risk prediction. A straightforward accuracy measurement is the proportion of correct predictions (correct rate, from 0 to 1) indicated by (6)

$$\text{Prediction accuracy} = \frac{\sum_{i=1}^{n} I(y_i = \hat{y}_i)}{n} \tag{6}$$

where $n$ is the data size, $y_i$ and $\hat{y}_i$ are true and model risk, respectively. The $I()$ function is the identify function defined in (7)

$$I(X) = \begin{cases} 1 & if\ X = True \\ 0 & if\ X = False \end{cases} \tag{7}$$

In this study, the primary objective is to build a model that can make predictions on data it has never seen before. To objectively measure the model's prediction accuracy, the data is randomly split into a training set and a test set in a training-test split. The model is trained on the training set of data (80%) and makes predictions on the response variable in the test set (20%) (Figure 1). The training-test split allows the model training and prediction to be performed on two different sets. The accuracy calculated on the test set is an unbiased estimation of model prediction accuracy.

Although the accuracy estimation is objective on the test set, it contains two prominent drawbacks. First, the split is fixed, excluding randomness. Second, the test data is wasted in model training due to the reduction of effective sample size. This study uses cross-validation to solve these two problems. Cross-validation randomly splits the data into $K$ equal-sized parts, trains the model on parts $K - 1$, and forces the model to predict the remainder (test part) of the data. This allows for repetition within the model or for every part of data to be treated as the test set. The final model prediction accuracy is obtained by averaging the accuracy of each part (Figure 2). We set $K$ to 5, train the model on four equally sized groups, and predict one left-out group. We use the R programming language and RStudio integrated development environment (IDE) to implement computations in this study [10,11].

**RESULTS**

The model provides a three-class prediction for each observation and total prediction accuracy. A model without any predictive power will randomly assign one of three classes, suggesting a 33% chance of correct class prediction. The model produced in this study increases prediction accuracy over the 33% baseline, shown by both training-test splits and cross-validation. The model prediction accuracy is 74% when evaluated by the training-test split. The model prediction accuracy is 80% when evaluated by the five-fold cross-validation. The 80% prediction accuracy generated suggests the model correctly predicted the TdP risk for 89 out of 112 observations, which indicates that the prediction accuracy is significantly increased by the model produced.

From the statistical perspective, the previous prediction result is a random variable. Its randomness originates from our data being a random sample from an unknown population (Figure 3). This randomness may never be eliminated because the population is not given. We need to obtain the prediction accuracy distribution to reveal the prediction uncertainty. To achieve this goal, we utilize the bootstrap resampling [12] to treat the current dataset as a population and sample with replacement from the dataset to imitate new samples. Each new bootstrap sample has the same sample size as the original data. Bootstrap provides a route to circumvent collecting different datasets by imitating the production of new samples from a population. For each training-test split in the cross-validation, there is a resample with replacement on the training set to generate a bootstrap training set. Then we predict the test set by the model trained on the bootstrap training set and save the prediction accuracy. We repeat this resampling process 1000 times and construct an empirical distribution for prediction accuracy across all test sets (Figure 4). The 95% confidence interval of prediction accuracy is (77%, 94%), which provides a range of values that contain the true prediction with up to a 95% probability. The range of this confidence interval is far greater than a 33% random guess.

The predictions obtained from the model are further dependent on individual predictors (variables). Permutation predictor importance tells which predictor contributes most to the complete prediction results [13]. We first train the model $m$ on dataset $D$ to obtain the prediction accuracy $acc$. Then each predictor $j$ is randomly shuffled in the test set (left-out drug). The model $m$ is applied on shuffled dataset $D_s$ and we calculate its prediction accuracy. The importance of predictor $j$ is calculated as the decrease of the prediction accuracy due to the random shuffling, as shown in Figure 5. Table 3 summarizes the top 5 most important variables. We normalize Table 3 by dividing the permutation predictor importance of the most important variable – ratio of the QTE interval. The other four most important variables are interval between the J point and the peak of T wave, interval between Q point and the end of T wave, interval between the J point and the peak of T wave, and ratio of TPE interval to QT interval.

# CONCLUSION

This study's results indicate that the model successfully uses data from *in vitro* preclinical studies to predict drug TdP risk. We train multinomial logistic regression on the isolated arterially perfused rabbit ventricular wedge preparation dataset. The variables found to be most influential by established literature are used in this study to generate a model to predict possible drug risk. Then an unbiased estimate of model performance is measured by training-test split and cross-validation. We also use bootstrap to calculate the model prediction uncertainty by reporting a 95% accuracy confidence interval. Finally, we compute the permutation predictor importance and interpret the model prediction. This study takes an exhaustive examination of drug TdP risk prediction; however, further analysis may be pursued. Potential future direction may include an in-depth review of important predictors. New *in vitro* preclinical studies may produce results different from ones previously generated with great regard to permutation predictor importance results. Understanding drug TdP risk with elevated prediction certainty may prove useful for drug regulation and further pharmaceutical research.

# ACKNOWLEDGEMENT

The authors would like to thank Dr. Gan-Xin Yan for his permission to use the wedge data from his lab.

# CONFLICT OF INTEREST

# Tables and Figures

| Variable |
| --- |
| TdP score proposed in the original study[6] |
| Interval between the J point and the start of T wave (JT) |
| Interval between the J point and the peak of T wave (JTP) |
| Ratio of the JTP interval at concentration $\geq 0$ to concentration $= 0$ |
| QS interval, measured at pace rate 2000 |
| QS interval, measured at pace rate 500 |
| Ratio of the QS interval at concentration $\geq 0$ to concentration $= 0$, measured at pace rate 500 |
| Interval between Q point and the end of T wave (QTE) |
| Ratio of the QTE interval at concentration $\geq 0$ to concentration $= 0$ |
| Ratio of QT interval to QS interval |
| Ratio of the interval between the peak and the end of T wave (TPE) to QT interval |
| Ratio of TPE interval[14] to QT interval |
| TPE Interval |
| Ratio of TPE interval at concentration $\geq 0$ to concentration $= 0$ |
| Score as a function of EAD[6] |

**Table 1.** Definition of variables in the dataset.

| High TdP Risk (8) | Intermediate TdP Risk (11) | Low TdP Risk (9) |
|---|---|---|
| Azimilide, Bepridil Disopyramide, Dofetilide Ibutilide, Quinidine Sotalol, Vandetanib | Astemizole, Chlorpromazine Cisapride, Clarithromycin Clozapine, Domperidone, Droperidol, Ondansetron, Pimozide, Risperidone, Terfenadine | Diltiazem, Loratadine Ranolazine, Metoprolol Mexiletine, Nifedipine Nitrendpine, Tamoxifen Verapamil |

**Table 2.** Drug list and associated risk.

| Variable | Permutation predictor importance (normalized) |
|---|---|
| Ratio of the QTE interval at concentration $\geq 0$ to concentration $= 0$ | 1.00 |
| Interval between the J point and the peak of T wave (JTP) | 0.81 |
| Interval between Q point and the end of T wave (QTE) | 0.72 |
| Interval between the J point and the start of T wave (JT) | 0.63 |
| Ratio of TPE interval to QT interval | 0.63 |

**Table 3.** The five variables with highest permutation predictor importance.

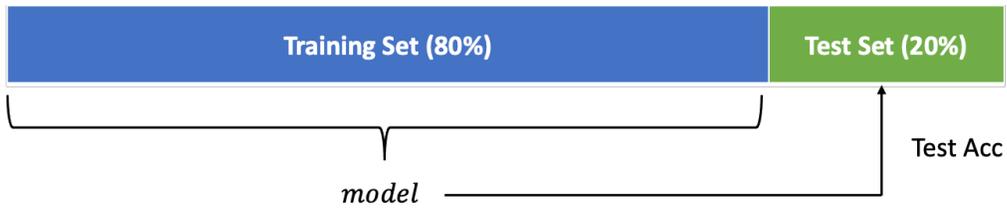

**Figure 1**. The measure of prediction accuracy by a training-test split.

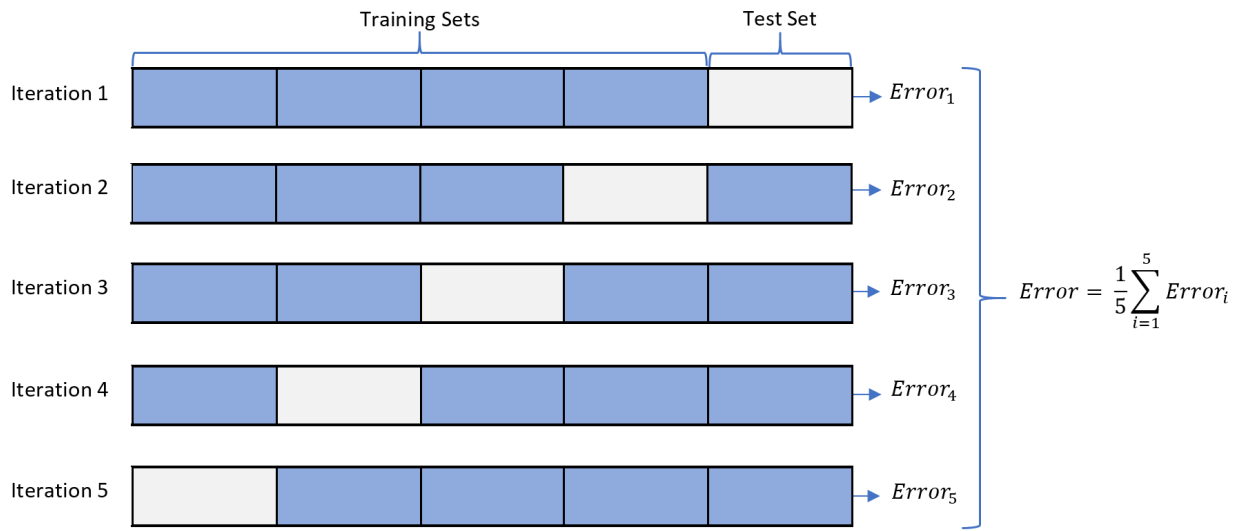

**Figure 2**. The measure of prediction accuracy (error) by 5-fold cross-validation.

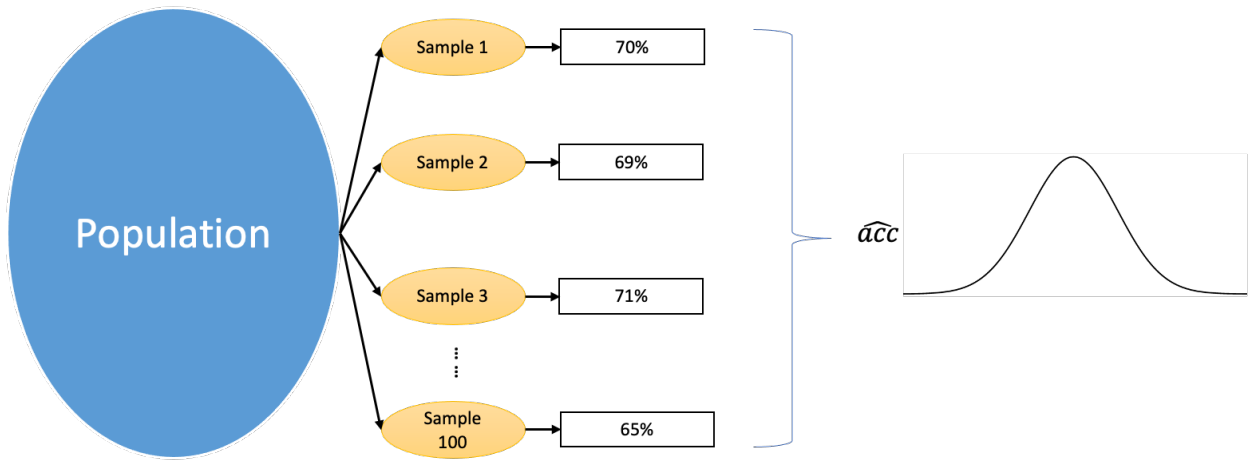

**Figure 3**. The demonstration of prediction randomness originating from a random sample.

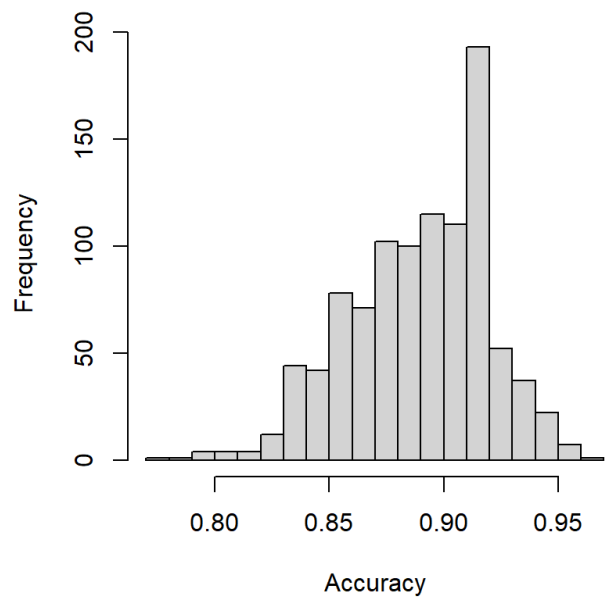

**Figure 4**. The empirical distribution of bootstrap prediction accuracy.

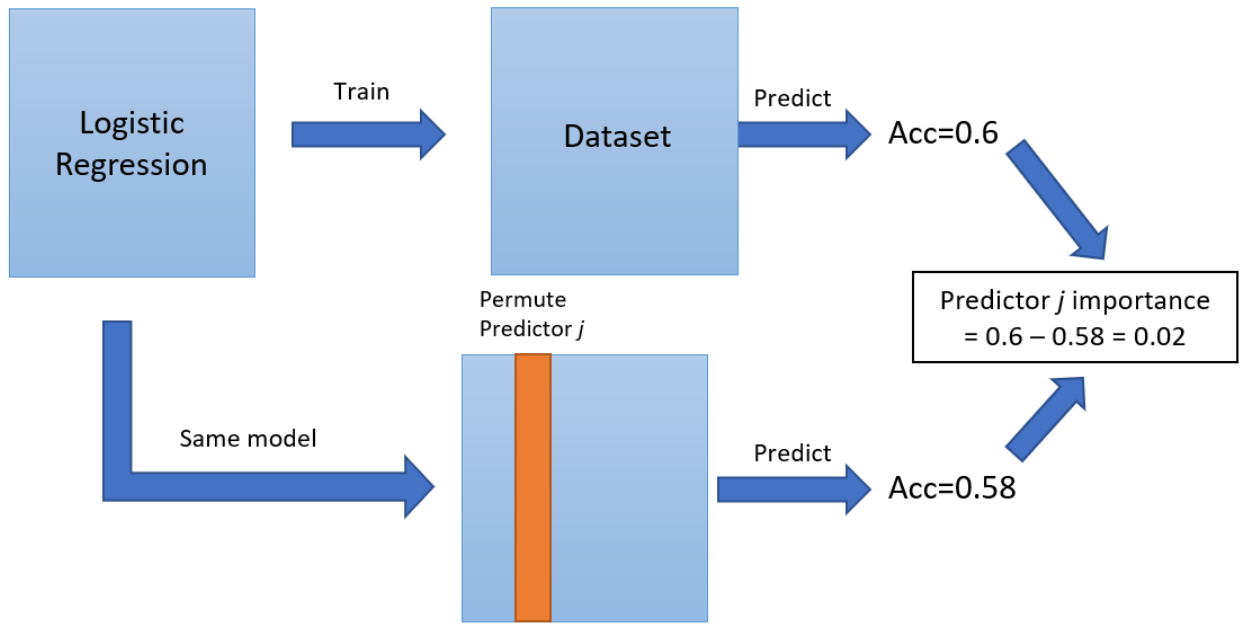

**Figure 5**. The calculation of permutation predictor importance.

# REFERENCE

1.  Roden, D. M. Drug-Induced Prolongation of the QT Interval. *N. Engl. J. Med.* **350**, 1013–1022 (2004).

2.  Xi, N. M., Hsu, Y.-Y., Dang, Q. & Huang, D. P. Statistical learning in preclinical drug proarrhythmic assessment. *J. Biopharm. Stat.* **32**, 450–473 (2022).

3.  Blinova, K. *et al.* International multisite study of human-induced pluripotent stem cell-derived cardiomyocytes for drug proarrhythmic potential assessment. *Cell Rep.* **24**, 3582–3592 (2018).

4.  Cavero, I. & Crumb, W. ICH S7B draft guideline on the non-clinical strategy for testing delayed cardiac repolarisation risk of drugs: a critical analysis. *Expert Opin. Drug Saf.* **4**, 509–530 (2005).

5.  Colatsky, T. *et al.* The Comprehensive in Vitro Proarrhythmia Assay (CiPA) initiative - Update on progress. *J. Pharmacol. Toxicol. Methods* **81**, 15–20 (2016).

6.  Liu, T. *et al.* Utility of Normalized TdP Score System in drug proarrhythmic potential assessment: A blinded in vitro study of CiPA drugs. *Clin. Pharmacol. Ther.* **109**, 1606–1617 (2021).

7.  Joshi, A., Dimino, T., Vohra, Y., Cui, C. & Yan, G.-X. Preclinical strategies to assess QT liability and torsadogenic potential of new drugs: the role of experimental models. *J. Electrocardiol.* **37 Suppl**, 7–14 (2004).

8.  James, G., Witten, D., Hastie, T. & Tibshirani, R. *An introduction to statistical learning*. (Springer, 2021).

9. *The elements of statistical learning: Data mining, inference, and prediction, second edition*. (Springer, 2009).

10. Core Team, R. & Others. R: a language and environment for statistical computing. *R Foundation for statistical computing, Vienna* (2013).

11. RStudio Team (2015) RStudio Integrated Development for R. *https://www.scirp.org › reference › ReferencesPapershttps://www.scirp.org › reference › ReferencesPapers*.

12. Efron, B. Bootstrap Methods: Another Look at the Jackknife. *aos* **7**, 1–26 (1979).

13. Breiman, L. Random Forests. *Mach. Learn.* **45**, 5–32 (2001).

14. Christophe, B. In silico study of transmural dispersion of repolarization in non-failing human ventricular myocytes: Contribution to cardiac safety pharmacology. *Br. J. Pharm. Res.* **7**, 88–101 (2015).